\documentclass[fleqn,10pt]{wlscirep}
\usepackage[utf8]{inputenc}
\usepackage[T1]{fontenc}
\usepackage{verbatim,graphicx,color,bbm,mathrsfs,csquotes}
\newcommand{\bra}[1]{\langle{#1} |}
\newcommand{\ket}[1]{|{#1}\rangle}
\newcommand{\braket}[2]{\langle{#1}| {#2}\rangle}
\newcommand{\ketbra}[2]{\vert {#1} \rangle \langle{#2}\vert}

\newcommand{\tcblue}[1]{\textcolor[rgb]{0,0,0.7}{#1}}

\newcommand{\tcer}[1]{\textcolor[rgb]{0,0,0}{#1}}

\title{Ultrastrong coupling probed by Coherent Population Transfer}
\author[1,2,3,+,*]{G. Falci}
\author[1,3,4,+]{A. Ridolfo}
\author[1,+]{P.G. Di Stefano}
\author[1,2,3,+]{E. Paladino}
\affil[1]{Dipartimento di Fisica e Astronomia "Ettore Majorana",
Universit\`a di Catania, Via Santa Sofia 64, 95123 Catania, Italy}
\affil[2]{CNR-IMM,  
UOS Universit\`a (MATIS), 
Via Santa Sofia 64, 95123 Catania, Italy.}
\affil[3]{INFN Sezione di Catania, Via Santa Sofia 64, 95123 Catania, Italy.}
\affil[4]{RIKEN, Theoretical Quantum Physics Laboratory, Saitama 351-0198, Japan}
\affil[*]{gfalci@dmfci.unict.it}
\affil[+]{these authors contributed equally to this work}

\begin{abstract}
Light-matter interaction, and the understanding of the fundamental physics behind, is the scenario of 
emerging quantum technologies. Solid state devices allow the exploration of new regimes where ultrastrong coupling strengths are comparable to subsystem energies, and new exotic phenomena like quantum phase transitions and ground-state entanglement occur. 
While experiments so far provided only spectroscopic evidence of ultrastrong coupling, we propose a new {\em dynamical} protocol for detecting virtual photon pairs in the dressed eigenstates. This is the fingerprint of the violated conservation of the number of excitations, which heralds the symmetry broken by ultrastrong coupling. We show that in flux-based superconducting architectures this photon production channel can be coherently amplified by Stimulated Raman Adiabatic Passage, providing a {\em unique} tool for an unambiguous dynamical detection of ultrastrong coupling in present day hardware. This protocol could be a benchmark for control of the dynamics of ultrastrong coupling  architectures, in view of applications to quantum information and microwave quantum photonics.
\end{abstract}
\begin{document}
\flushbottom
\maketitle
%
%
\thispagestyle{empty}


\section*{Introduction}
Strong coupling between atoms and quantized modes of an electromagnetic cavity~\cite{kb:206-haroche-raimond} provides a fundamental design building block of architectures for quantum technologies~\cite{ka:217-mohsenimartinis-nature-qtech}.
This regime is achieved when the coupling constant $g$ is large enough to overcome the individual decoherence rates of the mode and of the atom, $g\ll \kappa, \gamma$, and it has been observed in many experimental platforms from standard quantum optical 
systems~\cite{kb:206-haroche-raimond,kr:215-reisererrempe-rmp-qnets}, to architectures of 
artificial atoms (AA)~\cite{ka:199-imamoglu-prl-qdotJC,ka:204-wallraff-superqubit,kr:208-schoelkopf-nature-wiring}. In such systems small cavity volumes and large AA's dipoles yield values of $g$ up to $1\%$ of the cavity angular frequency $\omega_c$ and of the AA excitation energy $\varepsilon$. This allows to perform the rotating wave approximation (RWA) yielding the Jaynes-Cummings (JC) model of quantum optics~\cite{kb:206-haroche-raimond}, which describes the dynamics in terms of {\em individual} excitations exchanged between atom and mode. This process has been largely exploited for quantum control of AA-cavity architectures~\cite{ka:203-plastinafalci-prb,ka:204-wallraff-superqubit,kr:217-nori-review-supercqed}. Recently, fabrication techniques have allowed to go beyond, entering the regime of ultrastrong coupling (USC)~\cite{ka:205-ciuti-prb-intersubbandpolariton}, where $g \sim \omega_c, \varepsilon$ and the RWA breaks down. So far USC 
has been detected in superconducting~\cite{ka:210-niemczyck-natphys-ultrastrong,ka:210-diazmooji-prl-blochsiegert,ka:217-forndiazlupascu-natphys-usctunableflux,ka:217-yoshiarasemba-natphys-dscflux,ka:217-bosmansteele-multimodeusc} and semiconducting~\cite{ka:209-anapparabeltram-prb-ustr,ka:209-guentertredicucci-nature-switcusc,ka:212-scalari-science-USCTHz,ka:217-bayerlange-nanolett-dscsemicond} based architectures essentially via spectroscopic signatures. New physical processes emerge in the USC regime involving multiple photons and many qubits at once~\cite{kr:217-nori-review-supercqed}. Dynamical detection of population
transfer via a USC-specific channel (photon release by decay of the dressed ground state) has been proposed, using spontaneous emission pumping (SEP)~\cite{ka:213-stassisavasta-prl-USCSEP,ka:217-distefanosavastanori-njp-stimemission} or Raman oscillations~\cite{ka:214-huanglaw-pra-uscraman}. 
Several dynamical effects have also been predicted, from nonclassical photon statistics~\cite{ka:212-ridolfo-prl-photonblock,kr:217-nori-review-supercqed} to Casimir effect~\cite{ka:205-ciuti-prb-intersubbandpolariton,ka:207-deliberato-prl-uscdynamics} but despite the large interest, control in time is still an open experimental challenge.  
{Here we show that {\em coherent} dynamics amplifies fingerprints of USC in {\em available hardware}. 
Specifically we prove that a protocol similar to Stimulated Raman Adiabatic Passage
(STIRAP) in atomic physics~\cite{kr:198-bergmann-rmp-stirap,kr:201-vitanov-advatmolopt,kr:217-vitanovbergmann-rmp} operated in the so called Vee ($V$) configuration, provides a unique way to attain coherent population transfer via the USC channel.}
Demonstration of coherent dynamics in the USC regime would be a benchmark for quantum control, with appealing applications ranging from microwave quantum technologies~\cite{ka:210-ashhabnori-pra-uscdyn,ka:212-romerosolano-prl-ultrafastgates,ka:215-felicettisolano-scirept-stateengineernig-USC,ka:215-kyawsolano-prb-QECandUSC,ka:217-andersenblais-njp-transmonUSC} to dynamical control of quantum phase transitions~\cite{ka:213-ashab-pra-superradiance,ka:216-jaakorabl-pra-beyonddicke}.

\section*{Results}
\subsection*{The quantum Rabi model and STIRAP}
\begin{figure}[t!]
\centering
\resizebox{0.75\columnwidth}{!}{\includegraphics{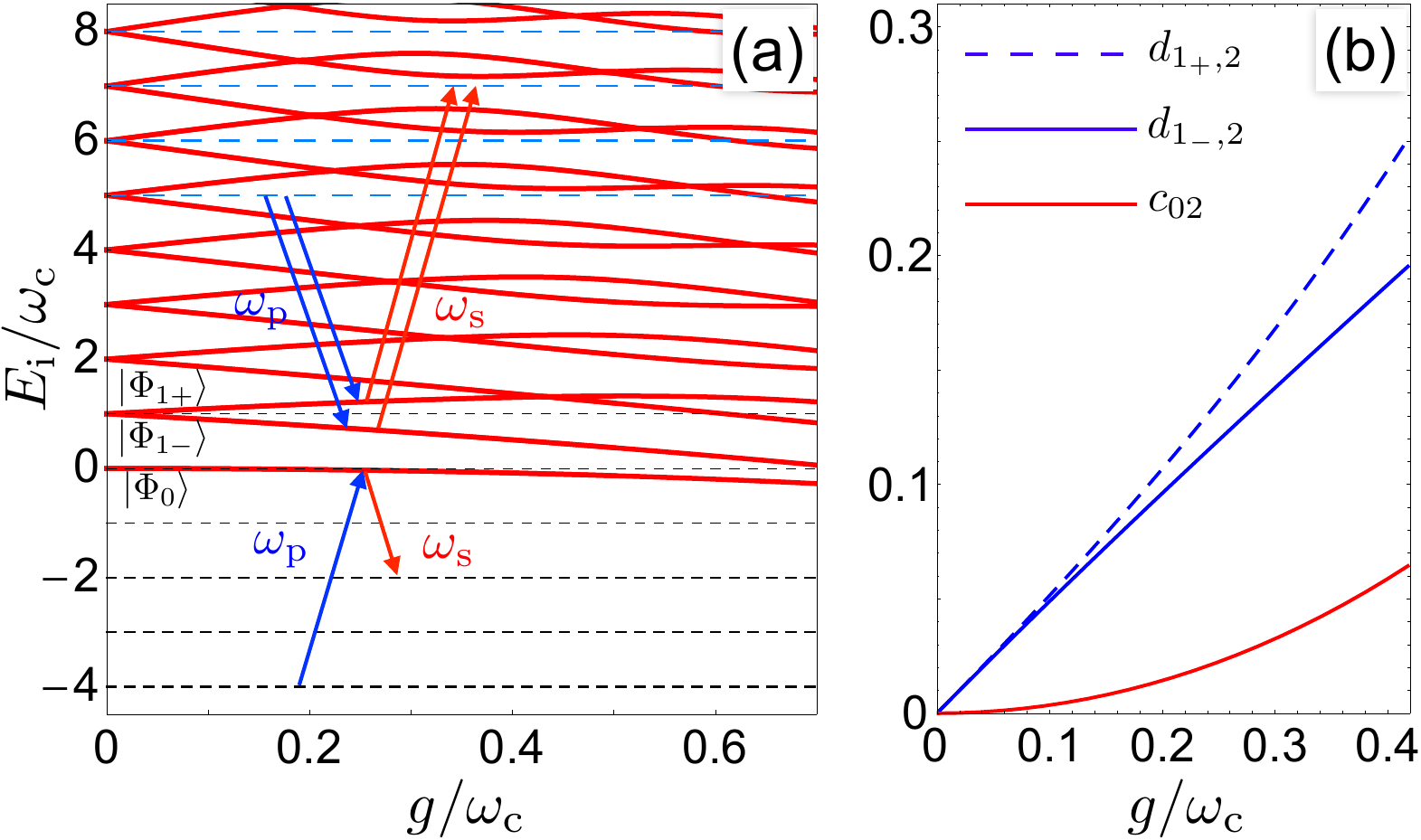}}
\caption{(color online) (a) Spectrum of $H_R$ Eq.(\ref{eq:rabiH2}) at resonance
(thick red lines): in the JC regime energies are linear in $g$; deviations yield the Bloch-Siegert shift, marking the onset of USC. Short-dashed lines are the energies of the eigenstates $\ket{nu}$ of the Hamiltonian $H_0$ ($\varepsilon^\prime = 4 \omega_c$): in the $\Lambda$ scheme two such states are coupled to $\ket{\Phi_0}$ by a pump ($\omega_p$) and a Stokes ($\omega_s$) laser. 
Long dashed lines are energies of $\ket{n u}$ for the $V$ scheme Eq.(\ref{eq:rabiH3-vee}). 
(b) Amplitudes $c_{02}(g) = \braket{2g}{\Phi_0}\propto g^2$ and $d_{1\pm,2}=\braket{2\,e}{\Phi_{1\pm}}\propto g$,
relevant for $\Lambda$ and $V$-STIRAP, respectively.
Here eigenstates $\{\ket{\Phi_0},\ket{\Phi_{N \pm}}\}$ of the Rabi model for small $g/\omega_c$ are labeled by the same quantum numbers of the JC model (see Methods~\cite{km:supplemental}).
\label{fig:usc1}}
\end{figure}

USC between a two-level atom (states $\{\ket{g},\ket{e}\}$ and energy spitting
$\varepsilon$), and a quantized harmonic mode is described by the quantum Rabi model
\begin{equation}
\label{eq:rabiH2}
H_R =  \varepsilon \, \ketbra{e}{e} + \omega_c  \,a^\dagger a
+ g \big(a^\dagger \ketbra{g}{e} + a \ketbra{e}{g}\big)
+ g  \big(a \ketbra{g}{e} + a^\dagger \ketbra{e}{g}\big)
\end{equation} 
$a$ ($a^\dagger$) being the annihilation (creation) operators acting on the oscillator
Hilbert space spanned by Fock states $\ket{n}$. The RWA  can be performed if  
$g,|\varepsilon-\omega_c| \ll \varepsilon,\omega_c$:   
the last \enquote{counterrotating} term is neglected yielding the JC Hamiltonian~\cite{kb:206-haroche-raimond,km:supplemental}, whose 
eigenstates $\ket{\phi_{N \pm}}$ have a defined number $N$ of excitations.
In the USC regime $g/\omega_c \sim 0.1-1$, the full $H_R$ comes into play, leading  to 
spectroscopic signatures (see Fig.~\ref{fig:usc1}(a)) as the Bloch-Siegert shift
observed in Ref.~\cite{ka:210-diazmooji-prl-blochsiegert}, 
and drastically altering the JC eigenstates which are mixed by USC.  
Eigenstates with energy $E_j$ of $H_R$ have the form 
$\ket{\Phi_j} = \sum_{n=0}^\infty [c_{j\,n} \ket{ng} + d_{jn} \ket{n\,e}]$, where the only symmetry 
left implies conservation of the parity of $N$.
In particular the ground state $\ket{\Phi_0}$, which in the JC model is 
factorized in the zero photon state and the atomic ground state $\ket{0\,g}$, 
acquires components with a finite number of photons, corresponding to nonvanishing $c_{0n}$ for $n$ even and $d_{0n}$ for $n$ odd. Proposals of dynamical detection of USC~\cite{ka:213-stassisavasta-prl-USCSEP,ka:214-huanglaw-pra-uscraman} aim at the detection of such virtual photons~\cite{ka:217-distefanosavastanori-njp-stimemission} by converting them to real ones.
To this end one considers a third ancillary atomic level $\ket{u}$ at a lower energy $-\varepsilon^\prime <0$. Assuming that the corresponding transitions are far detuned $\varepsilon^\prime \gg \omega_c$ and $\ket{u}$ is effectively uncoupled, 
{the Hamiltonian becomes~\cite{km:supplemental} } 
$H_0 = -\varepsilon^\prime \,\ketbra{u}{u} + H_R + \omega_c \,a^\dagger a \otimes \ketbra{u}{u}$.
In SEP~\cite{kr:201-vitanov-advatmolopt} population is pumped from $\ket{0u}$ to $\ket{\Phi_0}$ and may decay in $\ket{2u}$, due to the finite overlap $c_{02} := \braket{2g}{\Phi_0} \neq 0$. The process is forbidden in the JC limit, hence detection of this channel, uniquely leaving two photons in the mode, 
unveils USC~\cite{ka:213-stassisavasta-prl-USCSEP}. However SEP would have very low yield, since in most of the present implementations of USC architectures $c_{02}$ is not large enough. {This problem is overcome by a striking evolution of SEP, called $\Lambda$-STIRAP~\cite{kr:198-bergmann-rmp-stirap,kr:217-vitanovbergmann-rmp}, a remarkable technique in atomic physics recently extended to the solid state realm~\cite{ka:206-siebrafalci-optcomm-stirap,ka:208-weinori-prl-stirapqcomp,ka:209-siebrafalci-prb,ka:216-kumarparaoanu-natcomm-stirap,ka:216-xuhanzhao-natcomm-ladderstirap,ka:216-vepsalainen-photonics-squtrit}. Being based on quantum interference, STIRAP selectively addresses the target state with $\sim 100 \%$ efficiency. We now show how this allows to amplify coherently 
the USC channel.}
STIRAP is implemented by using a two-tone control field, $W(t)=\sum_{k=p,s}\mathscr{W}_k(t) \,\cos \omega_k t$, with slowly varying envelopes $\mathscr{W}_k(t)$ 
driving 
the AA~\cite{km:supplemental}. We assume for illustrative purposes that $\varepsilon^\prime  \gg \varepsilon$, and choose $\omega_{k} \sim \varepsilon^\prime$. Thus the field mainly couples to the two lowest atomic states, and the effective control Hamiltonian is given by 
\begin{equation}
\label{eq:lambda-control}
H_C(t) = W(t) \,(\ketbra{u}{g}+\ketbra{g}{u}) = 
W(t) \sum_{nj} \big[c_{jn} \ketbra{n\,u}{\Phi_j}+ c_{jn}^* \ketbra{\Phi_j}{n\,u}\big].
\end{equation}
By choosing $\omega_p \approx E_0 + \varepsilon^\prime$ and $\omega_s \approx E_0 + \varepsilon^\prime - 2 \omega_c$, 
and assuming moreover $\mathscr{W}_k,g \ll \omega_k$,  $H_C$ further simplifies 
yielding the $\Lambda$ scheme~\cite{kr:201-vitanov-advatmolopt} (see Fig.~\ref{fig:usc1}(a))
$$
H_C^\Lambda(t) = \frac{\Omega_{p}(t)}{2} \mathrm{e}^{i \omega_p t} \ketbra{0\,u}{\Phi_0} 
+  \frac{\Omega_{s}(t)}{2}   \mathrm{e}^{i \omega_s t} \ketbra{2\,u}{\Phi_0} + \mbox{h.c.}
$$
where $\Omega_p(t)=c_{00}(g) \mathscr{W}_p(t)$ and $\Omega_s(t)=c_{02}(g) \mathscr{W}_s(t)$ are 
the pump and the Stokes Rabi frequencies. 
Under the above assumptions the relevant dynamics involves three levels and it is described 
by~\cite{ka:217-falci-fortphys-fqmt} $H_3 = -\varepsilon^\prime \, \ketbra{0u}{0u} + (2 \omega_c-\varepsilon^\prime) \ketbra{2u}{2u} + E_0  \ketbra{\Phi_0}{\Phi_0} +H_C^\Lambda(t)$. For $\Lambda$-STIRAP\cite{kr:198-bergmann-rmp-stirap,kr:217-vitanovbergmann-rmp} 
the system is prepared in $\ket{0u}$ two pulses of width $T$ are shined in the ``counterintuitive'' sequence (the Stokes pulse {\em before} the pump pulse): this yields $~\sim 100\%$ population transfer to $\ket{2u}$, resulting from the adiabatic evolution of a \enquote{dark state}, stabilized by the destructive interference of the drives. Adiabaticity is attained using large pulse areas $\max_t [\Omega_k(t)] T > 10$ for both fields. Since $T$ is limited by the dephasing time  $T_\phi$, STIRAP requires appreciable USC mixing $c_{02}(g)$ to yield a large enough $\Omega_s$: if mixing is insufficient no adiabatic population transfer to $\ket{2u}$ occurs, 
whereas in the USC regime it occurs  with nearly unit probability. Therefore detection of $n=2$ photons in the 
cavity at the end of the protocol is a \enquote{smoking gun} for USC. 

This simple picture remains valid for the general multilevel dynamics, with driving fields coupled to all the allowed 
atomic transitions, Fig.~\ref{fig:lambda-failure}(a) showing that unit transfer probability is achieved. A key issue 
is that STIRAP requires $g$ large enough to guarantee adiabaticity for the Stokes pulse, 
$c_{02} \max_t[\mathscr{W}_s] T > 10$. This (soft) threshold depends {\em linearly} on $c_{02}(g)$, whereas 
the efficiency in SEP is much smaller, depending on $|c_{02}(g)|^2$ ($\propto g^4$ for small $g$, 
see Fig.~\ref{fig:usc1}(b)). Thus {\em coherence} in STIRAP amplifies population transfer by the USC channel.

Few remarks are in order. We have chosen equal peak Rabi frequencies $\max_t[\Omega_k]$, to ensure robustness against fluctuations, the property making STIRAP successful in atomic physics~\cite{kr:201-vitanov-advatmolopt,km:supplemental}, and checked that leakage from the three-level subspace is negligible, as expected since $\ket{\Phi_0}$ is not populated. 
For small $g$ STIRAP requires a large $\mathscr{W}_s$ (in our simulations 
its value would yield $e\!-\!g$ Rabi oscillations with $\Omega_0 :=600\,\mathrm{MHz}$) inducing dynamical Stark shifts, and producing a two-photon detuning $\delta(t)$ which may suppress population transfer~\cite{kr:217-vitanovbergmann-rmp,ka:216-distefano-pra-twoplusone}. The problem softens in the multilevel structure, where Stark shifts tend to self-compensate, and may be totally eliminated by using appropriately crafted control~\cite{ka:216-distefano-pra-twoplusone} (see Fig.~\ref{fig:lambda-failure}(a)).

\subsection*{Implementation}
\begin{figure}[t!]
\centering
\resizebox{0.88\columnwidth}{!}{\includegraphics{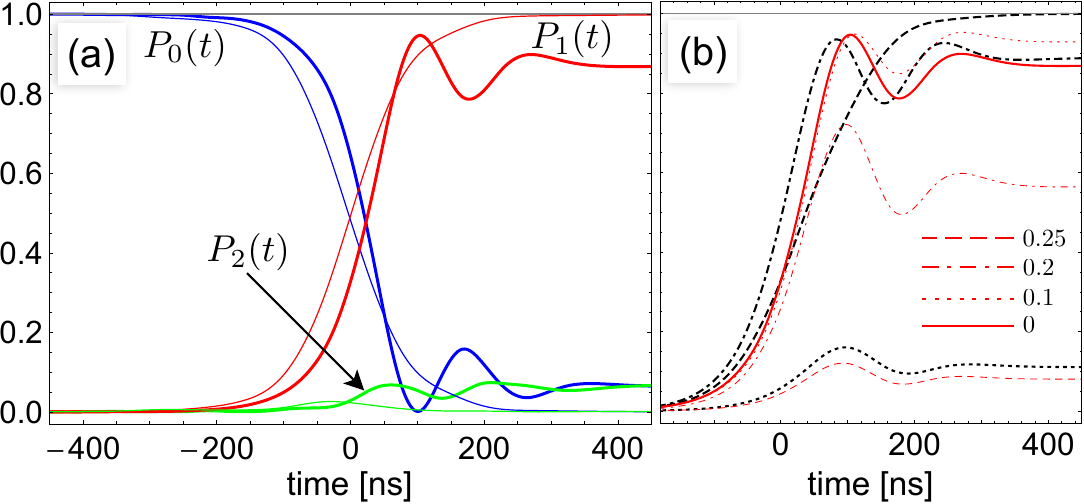}}
\caption{(color online) 
The full $\Lambda$-STIRAP dynamics of a AA-harmonic mode system at $\varepsilon=\omega_c$,
is studied, using 19 states,
$\varepsilon^\prime = 4 \varepsilon$ ($\alpha = 3$) and \tcer{$\Omega_0T= 900$}. Drives 
are coupled in the experimentally relevant \enquote{ladder} configuration 
$H_C(t)= W(t) [(\ketbra{u}{g} + (1/\eta) \ketbra{g}{e}) + \mbox{h.c.} ]$.
Population histories of the three relevant eigenstates 
$P_{0}(t) \leftrightarrow \ket{\Psi_{0u}}$,  $P_{1}(t) \leftrightarrow \ket{\Psi_{2u}}$ and 
$P_{2}(t) \leftrightarrow \ket{\Psi_0}$ are shown.
(a) Populations for $g=0.25$ and $g^\prime=0$, using Gaussian pulses (thick lines) and 
crafted pulses compensating Stark shifts (thin lines).
(b) Population $P_1(t)$ for $g=0$ and $g^\prime\neq 0$ in the RWA (thick black lines)
showing that the JC channel alone may led to population transfer.  
Red curves refer to $g=0.25$: for nonvanishing $g^\prime/\omega_c=0.1,0.2,0.25$
(corresponding to  $A=13.3,\,6.6,\,5.3$)
population transfer occurs due to USC only in the first case, the stray channel 
interfering destructively for larger $g^\prime$. 
\label{fig:lambda-failure}
}
\end{figure}

Implementation of the $\Lambda$ scheme in real devices faces two major problems. Anticipating the central result of our work. we claim that they can be overcome in a unique way by using STIRAP in the Vee ($V$) configuration. The first problem is the  
reliable detection scheme for the two-photons left in the cavity, which is problematic for THz-photons in semiconductors,
while GHz-photons in superconducting AA architectures can be detected with circuit-QED measurement technology~\cite{kr:217-nori-review-supercqed}. 
Thus multilevel superconducting AAs offer a natural implementation of our proposal. 
The second problem is the stray (dispersive) coupling of the mode to AA's transitions involving $\ket{u}$. 
This has a drastic impact on the reliability of protocols in $\Lambda$ configuration.
We discuss this point considering an additional stray coupling 
$g^\prime= \eta g$ between the mode and the AA $u\!-\!g$ transition.
Insight is gained by perturbation theory in $g^\prime$: the intermediate state 
$\ket{\Phi_0} \to \ket{\Psi_0}$ acquires a component onto $\ket{1u}$ and $\ket{2u}\to \ket{\Psi_{2u}}$ 
acquires a component onto $\ket{1g}$. Thus the \enquote{dipole} coupling to the Stokes field is modified: keeping only the corotating $g^\prime$ term, to lowest order we find
\begin{equation}
\label{eq:stokes-stray-lambda}
\Omega_s(t) \approx \Big[c_{02} - 
\frac{\sqrt{2} g^{\prime\, 2}}{(\varepsilon^\prime - \omega_c)^2 - g^2} \Big] \mathscr{W}_s(t) \;.
\end{equation}  
showing that $g^\prime$ opens a new channel  {\em already in the RWA}, which 
allows population transfer to $\ket{\Psi_{2u}}$ even if $c_{02}=0$. Therefore the final detection of two photons  
is {\em not any more} a \enquote{smoking gun} for USC. In general the 
stray coupling $g^\prime$ interferes destructively with the $g$-USC channel (see Fig.~\ref{fig:lambda-failure}(b)). 
STIRAP probes {\em selectively} the USC channel if and only if the correction in Eq.(\ref{eq:stokes-stray-lambda}) is so small that a large enough $T$ can be chosen, allowing adiabatic population transfer by the USC channel only. We obtain a necessary condition by treating in perturbation theory also the counterrotating $g$ (details are discussed later)
\begin{equation}
\label{eq:selective}
A := \frac{1}{2 \eta^2} \left |\frac{\alpha^2 - (g/\varepsilon)^2}{2 - (g/\varepsilon)^2} 
\right| \gtrsim 10
\end{equation} 
where $\alpha:=\varepsilon^\prime/\epsilon - 1$ is the anharmonicity of the AA spectrum. 
In this regime the two competing contributions to $\Omega_s(t)$ Eq.(\ref{eq:stokes-stray-lambda})
are both $\propto g^2$, thus the condition Eq.(\ref{eq:selective}) 
can be severe, and indeed {\em it is not met by any available design of superconducting AA}. In fact
in architectures based on the \enquote{flux} qubit exhibiting the largest figures of USC~\cite{ka:210-niemczyck-natphys-ultrastrong,ka:210-diazmooji-prl-blochsiegert,ka:217-forndiazlupascu-natphys-usctunableflux,ka:217-yoshiarasemba-natphys-dscflux}, 
selectivity is lost because the stray coupling is too strong, $\eta \gg 1$. 
The transmon desing~\cite{ka:217-bosmansteele-multimodeusc}, exhibiting the smallest decoherence rates~\cite{ka:212-rigettisteffen-prb-trasmonshapphire,kr:214-paladino-rmp}, offers smaller $\eta \approx 1/\sqrt{2}$, but  
long coherence times require small anharmonicity, $|\alpha| \lesssim 0.1$, and again selectivity is lost. Fig.~\ref{fig:lambda-failure} shows that it is not possible to select the USC channel even with parameters much more favorable than those of state-of the art devices. 
We remark that for the same reason all previous proposals of dynamical detection~\cite{ka:213-stassisavasta-prl-USCSEP,ka:214-huanglaw-pra-uscraman} are ruled out, i.e. present day hardware does not allow to detect unambiguously USC in the $\Lambda$ scheme.

\subsection*{Vee STIRAP scheme}
\begin{figure}[t!]
\centering
\resizebox{0.6\columnwidth}{!}{\includegraphics{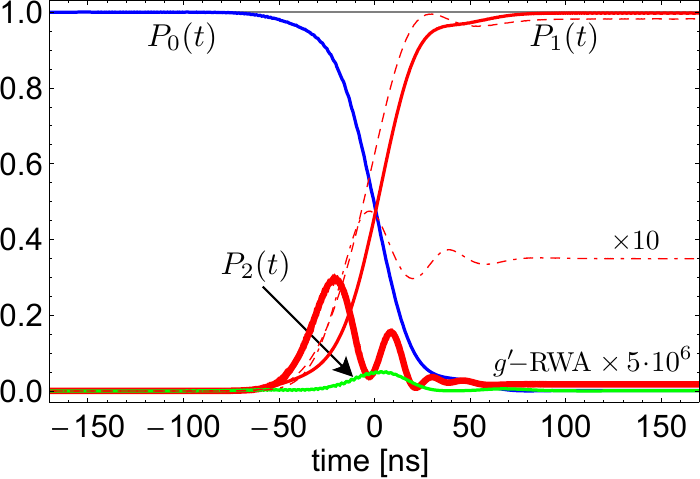}}
\caption{(color online) 
The full $V$-STIRAP dynamics via the intermediate state $\ket{\Phi_{1-}} \leftrightarrow P_2(t)$
is studied, using 26 states, $\varepsilon^\prime = 2.5 \ \varepsilon$ ($\alpha=1.5$), \tcer{$\Omega_0 T= 400$}. Population histories for $g=0.25\,\omega_c$ show that transfer to the desired target state occurs both 
in the absence (thin lines) and in the presence ($P_1(t)$, red dashed line) of a stray coupling $g^\prime= 2/3\,g$. On the contrary no population transfer occurs due to JC couplings: for  $g=0$ and $g^\prime=0.25\,\omega_c$ in the RWA, $P_1(t)\approx 0$ (red thick line). Instead the counterrotating $g^\prime$-USC term yields a small residual final population $P_1$ (red dash dotted line).
\label{fig:vee-control}
}
\end{figure}

The impasse is uniquely overcome by using the $V$-scheme for STIRAP. We consider a flux qubit, the lowest energy doublet being coupled to a harmonic mode in the USC regime~\cite{ka:210-niemczyck-natphys-ultrastrong,ka:210-diazmooji-prl-blochsiegert,ka:217-forndiazlupascu-natphys-usctunableflux,ka:217-yoshiarasemba-natphys-dscflux}, using the AA's second excited state 
as the ancillary $\ket{u}$. The system Hamiltonian is
\begin{equation}
\label{eq:rabiH3-vee}
H_0=H_{AA}+ H_1 + \omega_c \,a^\dagger a
\end{equation}
where 
$H_{AA} = \varepsilon \ketbra{e}{e} + (2+\alpha) \varepsilon \ketbra{u}{u}$ describes the 
flux AA, biased by an external magnetic flux $\Phi_x= \Phi_0/2$, 
$\Phi_0= h/2e$ being the flux quantum. This minimizes decoherence~\cite{kr:214-paladino-rmp,ka:203-paladino-advssp-decoherence} 
since $H_{AA}$ is symmetric with respect to 
fluctuations of $\Phi_x$. The corresponding selection rule forbids  
$g\!-\!u$ transitions, thus the full coupling to the mode reads 
$H_1 = g  \, (a+a^\dagger)  [(\ketbra{g}{e} + \eta \ketbra{e}{u}) + \mbox{h.c.}]$. 
We consider $\varepsilon=\omega_c$, and a general control field 
operated via the magnetic flux  
$H_C(t)= W(t) [(\ketbra{e}{u} + (1/\eta) \ketbra{e}{g}) + \mbox{h.c.} ]$.  
We exploit STIRAP  via one of the intermediate states $\ket{\Psi_{1\pm}}$, i.e. 
the two lowest excited states of the Hamiltonian (\ref{eq:rabiH3-vee}) 
We first neglect the stray coupling $e\!-\!u$ to the mode. Then 
$\ket{\Psi_{1\pm}}= \ket{\Phi_{1\pm}}$ (see Fig.\ref{fig:usc1}(a))
are eigenstates of $H_R$, with eigenvalues $E_{1\pm}$,   
reducing to the JC doublet $(\ket{0e} \pm \ket{1g})/\sqrt{2}$ 
when the counterrotating term is switched off. $V$-STIRAP
population transfer $\ket{0u} \to \ket{2u}$ is obtained by 
a two-tone $W(t)$, with 
$\omega_p = (1+\alpha) \varepsilon - E_{1\pm}$ and 
$\omega_s = (3+\alpha) \varepsilon - E_{1\pm}$. Insight is gained by projecting 
onto the three-level subspace $\mathrm{span}\{\ket{0u},\ket{2u},\ket{\Phi_{1\pm}}\}$, 
yielding an effective Hamiltonian with control
\begin{equation}
\label{eq:vee-control}
H_C^V(t)  = \frac{1}{2} \, \Big[ 
\Omega_{p}(t) \,\mathrm{e}^{-i \omega_p t} \,\ketbra{0\,u}{\Phi_{1\pm}} +
+ \Omega_{s}(t) \,\mathrm{e}^{-i \omega_s t} \,\ketbra{2\,u}{\Phi_{1\pm}} \Big] + 
\mbox{h.c.}
\end{equation}
where $\Omega_p(t) = d_{1\pm,0} \mathscr{W}_p(t)$ and $\Omega_s(t) = d_{1\pm,2} \mathscr{W}_s(t)$. 
Since in the absence of counterrotating terms $d_{1\pm,2} = \braket{2e}{\Phi_{1 \pm}}=0$
population transfer to $\ket{2u}$ can occur only in the USC regime. Indeed simulations considering the whole multilevel structure 
(Fig.~\ref{fig:vee-control}) show that  $\sim 100\%$ population transfer efficiency is achieved if and only if the USC regime 
is attained, {\em also when the stray coupling is present}. This striking success of $V$-STIRAP, while 
favored by large anharmonicity ($\alpha \ge 3$) and small ratio between the "ladder" 
matrix elements ($\eta \approx 1/3$) of flux-qubits, has a deeper and robust root: the stray JC coupling $g^\prime$ potentially spoiling USC-selectivity, {\em is not active in the $V$-scheme}. 
Indeed Fig.~\ref{fig:vee-control} shows that USC-selective population transfer is attained with  
parameters $\alpha = 1.5$ and $\eta=2/3$, not satisfying the requirement Eq.(\ref{eq:selective}), 
and even worse parameters work.  
Technical details on the JC channel suppression are given in the next subsection. Here we mention that leading corrections to the control Hamiltonian 
(\ref{eq:vee-control}) due to the corotating $g^\prime$ vanish because 
$\braket{nu}{\Psi_{1\pm}}=0$ at lowest order in perturbation theory. 
This makes  $V$-STIRAP unique as a \enquote{smoking gun} for dynamically probing  
USC, which again is witnessed by the detection of $n=2$ photons at the end of the 
protocol. The probability is approximately the population of $\ket{\Psi_{2u}}$, 
the stray $g^\prime$ determining only a small probability of detecting a different $n$  
($\sim [\eta \, g/(\alpha \varepsilon)]^2 \sim 10^{-1}(g/\omega_c)^2$, for the simulation in Fig.~\ref{fig:vee-control}).

The suppression of the JC channel makes $V$-STIRAP USC-selective also for smaller $\alpha$, 
thus lower microwave frequencies can be used for the driving fields, a key experimental advantage.
Another asset of $V$-STIRAP is that since $d_{1\pm\,2}(g) > c_{0\,2}(g)$ (see Fig.~\ref{fig:usc1}(b))
coupling with the Stokes field is larger. Therefore sufficient adiabaticity is attained with smaller $T$: this minimize decoherence effects and/or softens the problem of stray dynamical Stark shifts since weaker Stokes fields $\mathscr{W}_s$ can be used.
Notice indeed that in Fig.~\ref{fig:vee-control} shorter time scales than in Fig.~\ref{fig:lambda-failure} were used, and that Stark shifts are not apparent. 

Few remarks are in order. We took for granted preparation in the state $\ket{0u}$: for $g=0$, 
it is prepared from $\ket{0g}$ by standard pulse sequences~\cite{ka:215-peterergustaffson-prl-transmexclevel}. 
But if $g\neq0$ the ground state is $\ket{\Phi_0}$ and the above procedure
prepares a state which may contain photons. Since the probability
$\propto |c_{0n}(g)|^2$ is very small for not so large $g$, while $d_{1\pm\,2}(g)$ 
is large enough to guarantee $V$-STIRAP population transfer, we expect only some harmless 
lack of accuracy. Similar arguments ensure that for reasonable parametrization mixing 
of $\ket{0u}$ due to $g^\prime \neq 0$ is also small. In any case more accurate preparation protocols 
may be designed to minimize errors.

Concerning decoherence, 
we know that STIRAP is mainly sensitive to fluctuations in the $\mathrm{span}\{\ket{0u},\ket{2u}\}$ subspace and rather insensitive to other processes~\cite{ka:213-falci-prb-stirapcpb,kr:214-paladino-rmp}. Efficient population transfer requires $T < T_\phi$, where $1/T_\phi$ is the 
decoherence rate in the ``trapped`` subspace, which is approximately
the sum of the decay rate $\kappa$ of the mode and the decay rate $\gamma_{u \to e}$ of 
$\ket{u}$ in high-quality devices.
In such systems these rates are very small, allowing for $T$ up to 
several dozens of $\mu \mathrm{s}$. In devices used for USC spectroscopy the mode has a 
much smaller quality factor, but there should be no fundamental tradeoff between large $g$ and 
decoherence of the mode alone, allowing for the fabrication of devices exploiting the coherent 
dynamics in the USC regime. In alternative, with the standard design of high-quality devices
large effective couplings $g_{eff} \sim \sqrt{N} g$ could be attained by using few weakly coupled AAs. 
We checked the dynamics for $N=4$ AAs, and we reproduced results of Fig.(\ref{fig:vee-control}) using half of the value of $g$~\cite{ka:218-ridolfofalci-epj-usc}. 
We also checked that the protocol is robust against possible inhomogeneities of the individual couplings of AAs and the possible presence of stray additional modes at multiple frequencies. 

Finally we stress that for the detection of USC it would be sufficient to monitor the population of the Fock states 
$\ket{n\ge 2}$ during part of the protocol. Some transient population of the intermediate state is also tolerable, softening the adiabaticity requirement. Decoherence times $T_\phi \sim T$ can also be tolerated~\cite{ka:213-falci-prb-stirapcpb} since at worst efficiency of the USC-selective channel would be larger than 30\%. This opens perspectives also for semiconducting structures, where USC-selective $\Lambda$-STIRAP could be observed with some progress in techniques for detecting excess THz photons. 

\subsection{Effect of stray coupling}
\paragraph{Lambda scheme}
We now discuss in more detail the effect of stray couplings. In the $\Lambda$ scheme 
we add to the undriven Hamiltonian, Eq.~(\ref{eq:rabiH3}), the additional stray coupling 
$g^\prime= \eta g$ between the mode and the  $u-g$ transition, which is the relevant one for AAs. Specifically we consider the more general Hamiltonian 
$\tilde{H}_0 = \tilde{H}_{JC}+ \tilde{H}_c$ where
\begin{equation}
\label{eq:H-threelevel-lambda-JC}
\tilde{H}_{JC} = 
 -  \varepsilon^\prime \ketbra{u}{u} + \varepsilon \ketbra{e}{e} + \omega_c\,a^\dagger a 
+ 
\big[(g \, a \ketbra{e}{g} + g^\prime a \ketbra{g}{u}) + \mbox{h.c.}\big]
\end{equation}
contains all corotating couplings and 
\begin{equation}
\label{eq:H-threelevel-lambda-count} 
\tilde{H}_c = (g_c \, a^\dagger \ketbra{e}{g} + g^\prime_c a^\dagger \ketbra{g}{u}) + \mbox{h.c.}
\end{equation}
are the counterrotating terms. 
Again $\tilde{H}_{JC}$ conserves the number $N$ of 
excitations of the three-level atom plus the harmonic mode, 
$N \leftrightarrow a^\dagger a + \ketbra{g}{g} + 2\ketbra{e}{e}$. This determines the structure of its eigenstates,  denoted by $\ket{\psi_{j}}$. 
Eigenstates of $\tilde{H}_0$ do not possess a well defined $N$ and have the general structure
\begin{equation}
\label{eq:eigenstates3-factorbasis}
\ket{\Psi_j} = \sum_{n=0}^\infty c_{j,n} \ket{n g} + d_{j,n} \ket{n e} + f_{j,n} \ket{n u}
\end{equation}
Again the parity of $N$ is conserved thus many amplitudes vanish. If the counterrotating $\tilde{H}_c$ can be treated as a perturbation we can use for $\ket{\Psi_j}$ the same quantum numbers of the JC eigenstates, $j \equiv (N,\tau)$ (see Methods~\cite{km:supplemental}).  

To fix the ideas we consider $\omega_c = \varepsilon < \varepsilon^\prime$, 
and focus on the limit $g^\prime \ll |\epsilon^\prime - \omega_c|$, hereafter referred as the 
dispersive regime (for the stray coupling). 
In this limit it is convenient to classify $\ket{\Psi_j}$ with the same quantum numbers 
of the JC eigenstates for $g \neq 0$ and $g^\prime=0$, namely  
$\{\ket{nu} \} \bigcup \{\ket{\phi_0},\ket{\phi_{1 \mp}}, \dots\}$. Focusing on a simple picture where STIRAP involves only levels resonantly coupled by the drives, and letting the two tone pulse couple the intermediate state $\ket{0g} \to \ket{\Psi_{0}}$ with $\ket{\Psi_{0u}}$ (pump) and 
$\ket{\Psi_{2u}}$ (Stokes) the physics is described by an effective three-level Hamiltonian. We now need the matrix elements of the control field in this subspace.  To this end we must first diagonalize $\tilde{H}_0 = \tilde{H}_{JC}+ \tilde{H}_c$, Eqs.(\ref{eq:H-threelevel-lambda-JC},\ref{eq:H-threelevel-lambda-count}). The main structure of the amplitudes is captured by diagonalizing $\tilde{H}_0$ in the $12-$dimensional subspace spanned by the factorized states with $N \le 4$ excitations. This subspace is enough to account for counterrotating terms in leading (first) order, whereas corotating terms are treated exactly. For the pump field we find
$$
\braket{\Psi_{0}}{H_C| \Psi_{0u}} \approx W_p(t) \,[c^*_{0,0} f_{0u,0}+ f^*_{0,1} c_{0u,1}] 
$$
The leading term of the matrix element can be found by noticing that all the amplitudes are of order zero in the small quantity $g^\prime_c/(\epsilon^\prime + \omega_c)$ 
except $c_{0u,1}$ which is first order, and can be neglected. In particular since in the dispersive regime $f_{0u,0} \approx 1$ the resulting matrix element is $\approx c^*_{0,0}  W_p(t)$, i.e. in leading order in the stray coupling the pump matrix element in 
$H_C^\Lambda$ is unaffected. Instead for the Stokes field we find substantial differences. The matrix element in the   $12-$dimensional subspace is 
$$
\braket{\Psi_{0}}{H_C| \Psi_{2u}} \approx W_s(t) [c^*_{0,2}f_{2u,2}+
f^*_{0,1} c_{2u,1}+ c^*_{0,0}f_{2u,0}] 
$$
The first term is the Stokes matrix element in $H_C^\Lambda$ modified by the stray coupling $g^\prime$. 
Indeed in the dispersive regime we can approximate $f_{2u,2}\approx 1$, and we recover the expression in $H_C^\Lambda$. 
The two extra terms 
are due to the stray coupling only: the second term is due to the JC part
$g^\prime$, whereas the third depends on the counterrotating part and vanishes if  
$g^\prime_c=0$. 

These extra terms are important since in the physical case $g_c=g$ and $g^\prime_c = g^\prime$ they may be of the same order of the first one, modifying substantially the matrix element. 
To clarify this point we notice that in the dispersive regime for the stray coupling we also have $c^*_{0,0} \approx 1$ therefore 
$$
\braket{\Psi_{0}}{H_C(t)| \Psi_{2u}} \approx 
W_s(t) \,[c^*_{0,2}+ f^*_{0,1} c_{2u,1}+ f_{2u,0}]
\approx W_s(t)\,\big[c^*_{0,2}+ c_{2u,1} \big( \frac{g^\prime}{\epsilon^\prime-\omega_c} +
\frac{g_c^\prime}{2 \omega_c} \big) \big]
$$
where the last line is obtained by estimating extra terms by perturbation theory in $g^\prime$ and 
$g^\prime_c$, which gives in leading order 
$$
\begin{aligned}
f_{0,1} = \braket{1u}{\Psi_0} &\approx \frac{g^\prime}{\varepsilon^\prime -\omega_c}
\\
c_{2u,1}  = \braket{1 g}{\Psi_{2u}}  &\approx
- \sqrt{2} g^\prime \frac{\varepsilon^\prime - \omega_c}{ 
(\varepsilon^\prime - \omega_c)^2 - g^2}
\\
f_{2u,0} = \braket{0 u}{\Psi_{2u}}  &\approx - \frac{\sqrt{2} g_c^\prime g^\prime}{2 \omega_c} 
\frac{\varepsilon^\prime - \omega_c }{ 
(\varepsilon^\prime - \omega_c)^2 - g^2} \approx \frac{g_c^\prime }{ 2 \omega_c} \, c_{2u,1}
\end{aligned}
$$
Notice that our definitions imply that  $c^*_{0,2}$ and $c_{2u,1}$ have different signs therefore the 
extra terms due to stray coupling interfere destructively with the amplitude due to the 
counterrotating $g_c$ marking USC. In particular this happens even if the stray coupling is purely 
corotating, i.e. for $g^\prime\neq0$ and $g^\prime_c=0$. This remarkable illustrative case corresponds to Eq.~(\ref{eq:stokes-stray-lambda}), showing that a non-vanishing $g^\prime$ yields a nonzero Stokes matrix element even if $g_c=0$. This is sufficient to determine population transfer to the two-photon target state, if the matrix element the is large enough to guarantee the global adiabaticity condition for STIRAP, $\max_t |\Omega_k(t)| T > 10$. In  
Fig.~\ref{fig:lambda-failure}b (black curves) we show that this picture describes also 
the physics beyond perturbation theory (a larger system of 30 states is diagonalized)
and accounting for the general coupling structure of the two-tone driving field $W(t)$.

The necessary condition for detecting unambiguously the USC channel, Eq.(\ref{eq:selective})
is found by arguing that this 
latter must satisfy the global adiabaticity condition, $|c_{02}| \mathscr{W}_s T > 10$, while the stray channel does not $|f^*_{0,1} c_{2u,1}| \mathscr{W}_s T < 10$. This determines the weaker criterion $|c_{02}|/|f^*_{0,1} c_{2u,1}| \gg 10$. Eq.(\ref{eq:selective} is obtained by using the perturbative result (\ref{eq:rabi-tls-perturbative}) for $c_{0,2}$, and letting $g_c = g$.
Destructive interference of the $g_c=g$ and the $g^\prime$ channels is illustrated in 
Fig.~\ref{fig:lambda-failure}b  (red curves), where it is seen that this physical picture holds beyond perturbation theory.

\begin{figure}[t!]
\centering
\resizebox{0.35\columnwidth}{!}{\includegraphics{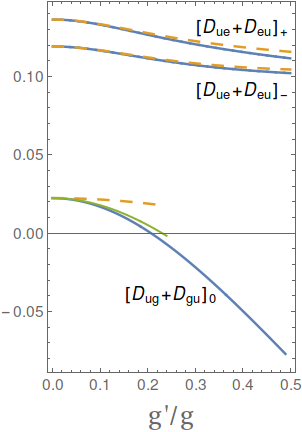}}
\caption{(color online) \enquote{Dipole} matrix elements for the Stokes field when the three-level system is coupled in ladder configuration with an harmonic mode, in the 
USC regime, $g=0.25 \,\omega_c$, with one mode at resonance $\epsilon = \omega_c$, 
as a function of the relative strength of the stray coupling $\eta=g^\prime/g$. 
Here $[D_{ug}+D_{gu}]_0 = \bra{\Psi_{2u}} \big[\ketbra{u}{g}+\ketbra{g}{u}\big] \ket{\Psi_0}$ is relevant for $\Lambda$-STIRAP. The full thick line is obtained by diagonalizing a large system, and it is identical to the result of the diagonalization in the 12-dimensional subspace containing $N\le 4$ excitations (not shown). The full thin line is the approximation we start with to derive 
Eq.~(\ref{eq:stokes-stray-lambda}) and Eq.(\ref{eq:selective}), and includes 
only the $g^\prime$ corotating interaction. The dashed line is obtained neglecting leading corrections in $g^\prime$. 
These results quantitatively illustrate the scenario on the failure of $\Lambda$-STIRAP. Instead the behavior of $[D_{ue}+D_{eu}]_\pm = \bra{\Psi_{2u}} \big[\ketbra{u}{e}+\ketbra{e}{u}\big] \ket{\Psi_{1 \pm}}$, which enters $V$-STIRAP, is weakly dependent on $g^\prime$, except for the wavefunction renormalization: the result obtained by diagonalizing a large system (full thick line) is identical to the result of the diagonalization in the subspace containing $N\le 6$ excitations (not shown), where terms due to the corotating $g^\prime$ are absent. 
Moreover leading corrections in the counterrotating $g^\prime$ are small. 
\label{fig:stokes-lambdavee}}
\end{figure}

\paragraph{Vee scheme}
For STIRAP in the Vee configuration the ancillary atomic level $\ket{u}$ has higher energy. Again we start from a more general Hamiltonian 
$\tilde{H}_0 = \tilde{H}_{JC}+ \tilde{H}_c$ where 
\begin{equation}
\label{eq:H-threelevel-vee-JC}
\tilde{H}_{JC} = 
 \varepsilon \, \ketbra{e}{e} + (\varepsilon+\varepsilon^\prime) \, \ketbra{u}{u} +
\omega_c  \,a^\dagger a + 
\big[(g \, a \ketbra{e}{g} + g^\prime a \ketbra{u}{e}) + \mbox{h.c.}\big]
\end{equation}
contains all corotating couplings, the counterrotating terms being
\begin{equation}
\label{eq:H-threelevel-vee-count} 
\tilde{H}_c = (g_c \, a^\dagger \ketbra{e}{g} + g^\prime_c a^\dagger \ketbra{u}{e}) + \mbox{h.c.}
\end{equation}
Eigenstates of $\tilde{H}_{JC}$ have a well defined number of excitations, 
$N \leftrightarrow a^\dagger a + \ketbra{e}{e} + 2\ketbra{u}{u}$.
They are the isolated ground state $\ket{\Phi_{0}}=\ket{0g}$, a single excitation $g$-JC doublet $\ket{\Phi_{1\mp}}=\ket{\phi_{1\mp}}$, and $N\ge 2$ triplets $\ket{\Phi_{N\tau}}$ as before. 
We are interested to the dispersive regime for the stray coupling $g^\prime \ll \epsilon^\prime - \omega_c$. As for $\Lambda$-STIRAP eigenstates are conveniently enumerated by the quantum numbers $\{0u,1u, \dots, 0, 1\mp,2\mp, \dots \}$, but the energy spectrum is different.

Eigenstates of $\tilde{H}_0$ have the same structure Eq.(\ref{eq:eigenstates3-factorbasis}), the counterrotating terms mixing subspaces with the same parity of $N$. Following the same steps of the analysis of $\Lambda$-STIRAP  we focus on the part of the control term 
$H_C(t) = W(t) [\ketbra{u}{e}+ \ketbra{e}{u}]$ resonant with the $\ket{\Psi_{0u}} \leftrightarrow \ket{\Psi_{1 \pm}}$ (pump) and the $\ket{\Psi_{2u}} \leftrightarrow \ket{\Psi_{1 \pm}}$ transitions calculating matrix elements by diagonalization of $\tilde{H}_0$ in the $18-$dimensional subspace spanned by states with up to $N \le 6$ excitations. For the pump field we find
$$
\braket{\Psi_{1 \pm}}{H_C| \Psi_{0u}} \approx W_p(t) \,
[d^*_{1 \pm,0} f_{0u,0}+ f^*_{1 \pm,1} d_{0u,1}] 
$$
which in leading order in $g_c$ and $g^\prime_c$ and in the dispersive regime reduces to 
$\approx W_p(t)\,d^*_{1 \pm,0}$. Therefore at this level of accuracy the stray coupling does not modify the pump matrix element in $H_C^V$ Eq.~(\ref{eq:vee-control}). For the Stokes field, in leading  order in $g_c$ and $g^\prime_c$, we find
$$
\braket{\Psi_{1 \pm }}{H_C| \Psi_{2u}} \approx 
W_s(t) [d^*_{1 \pm,2}f_{2u,2}+ d^*_{1 \pm,0}f_{2u,0}] 
$$
Here $d^*_{1 \pm,2}$ and $f_{2u,0}$ are first order in the small counterrotating couplings $g_c$ and $g_c^\prime$ whereas in the dispersive regime $f_{2u,2},d^*_{1 \pm,0} \to 1$. The remarkable fact is that the matrix element vanishes for both $g_c, g_c^\prime \to 0$, therefore to this level of accuracy no population transfer may occur due to the corotating stray coupling $g^\prime$. This is clearly shown in Fig.~\ref{fig:vee-control}, where the whole structure of the drive is accounted for, showing that  stray population transfer due to the corotating $g^\prime$ is suppressed by six orders of magnitude even at an accuracy level beyond perturbation theory. Extra contribution is possibly due to the stray counterrotating $g_c^\prime$, which in the physical situation of 
Fig.~\ref{fig:vee-control} is also very small. Summing up corotating couplings do not produce population transfer, whose observation  marks unambiguously the emergence of  USC. 

\begin{thebibliography}{10}
\urlstyle{rm}
\expandafter\ifx\csname url\endcsname\relax
  \def\url#1{\texttt{#1}}\fi
\expandafter\ifx\csname urlprefix\endcsname\relax\def\urlprefix{URL }\fi
\expandafter\ifx\csname doiprefix\endcsname\relax\def\doiprefix{DOI: }\fi
\providecommand{\bibinfo}[2]{#2}
\providecommand{\eprint}[2][]{\url{#2}}

\bibitem{kb:206-haroche-raimond}
\bibinfo{author}{Haroche, S.} \& \bibinfo{author}{Raimond, J.~M.}
\newblock \emph{\bibinfo{title}{Exploring the Quantum}}
  (\bibinfo{publisher}{Oxford University press}, \bibinfo{year}{2006}).

\bibitem{ka:217-mohsenimartinis-nature-qtech}
\bibinfo{author}{Mohseni, M.} \emph{et~al.}
\newblock \bibinfo{journal}{\bibinfo{title}{Commercialize quantum technologies in five years}}.
\newblock {\emph{\JournalTitle{Nature}}} \textbf{\bibinfo{volume}{543}},
  \bibinfo{pages}{171} (\bibinfo{year}{2017}).

\bibitem{kr:215-reisererrempe-rmp-qnets}
\bibinfo{author}{Reiserer, A.} \& \bibinfo{author}{Rempe, G.}
\newblock \bibinfo{journal}{\bibinfo{title}{Cavity-based quantum networks with
  single atoms and optical photons}}.
\newblock {\emph{\JournalTitle{Rev. Mod. Phys.}}}
  \textbf{\bibinfo{volume}{87}}, \bibinfo{pages}{1379--1418},
  \doiprefix\url{10.1103/RevModPhys.87.1379} (\bibinfo{year}{2015}).

\bibitem{ka:199-imamoglu-prl-qdotJC}
\bibinfo{author}{Imamoglu, A.} \emph{et~al.}
\newblock \bibinfo{journal}{\bibinfo{title}{Quantum information processing
  using quantum dot spins and cavity qed}}.
\newblock {\emph{\JournalTitle{Phys. Rev. Lett.}}}
  \textbf{\bibinfo{volume}{83}}, \bibinfo{pages}{4204--4207},
  \doiprefix\url{10.1103/PhysRevLett.83.4204} (\bibinfo{year}{1999}).

\bibitem{ka:204-wallraff-superqubit}
\bibinfo{author}{Wallraff, A.} \emph{et~al.}
\newblock \bibinfo{journal}{\bibinfo{title}{Strong coupling of a single photon
  to a superconducting qubit using circuit quantum electrodynamics}}.
\newblock {\emph{\JournalTitle{Nature}}} \textbf{\bibinfo{volume}{421}},
  \bibinfo{pages}{162--167}, \doiprefix\url{10.1038/nature02851}
  (\bibinfo{year}{2004}).

\bibitem{kr:208-schoelkopf-nature-wiring}
\bibinfo{author}{Schoelkopf, R.~J.} \& \bibinfo{author}{Girvin, S.~M.}
\newblock \bibinfo{journal}{\bibinfo{title}{Wiring up quantum systems}}.
\newblock {\emph{\JournalTitle{Nature}}} \textbf{\bibinfo{volume}{451}},
  \bibinfo{pages}{664} (\bibinfo{year}{2008}).

\bibitem{ka:203-plastinafalci-prb}
\bibinfo{author}{Plastina, F.} \& \bibinfo{author}{Falci, G.}
\newblock \bibinfo{journal}{\bibinfo{title}{Communicating josephson qubits}}.
\newblock {\emph{\JournalTitle{Phys. Rev. B}}} \textbf{\bibinfo{volume}{67}}
  (\bibinfo{year}{2003}).

\bibitem{kr:217-nori-review-supercqed}
\bibinfo{author}{Gu, X.}, \bibinfo{author}{Frisk~Kockum, A.},
  \bibinfo{author}{Miranowicz, A.}, \bibinfo{author}{Liu, Y.-X.} \&
  \bibinfo{author}{Nori, F.}
\newblock \bibinfo{journal}{\bibinfo{title}{Microwave photonics with
  superconducting quantum circuits}}.
\newblock {\emph{\JournalTitle{Physics Reports}}}
  \textbf{\bibinfo{volume}{718–719}}, \bibinfo{pages}{1--102}
  (\bibinfo{year}{2017}).
\newblock \bibinfo{note}{ArXiv:1707.02046}.

\bibitem{ka:205-ciuti-prb-intersubbandpolariton}
\bibinfo{author}{Ciuti, C.}, \bibinfo{author}{Bastard, G.} \&
  \bibinfo{author}{Carusotto, I.}
\newblock \bibinfo{journal}{\bibinfo{title}{Quantum vacuum properties of the
  intersubband cavity polariton field}}.
\newblock {\emph{\JournalTitle{Phys. Rev. B}}} \textbf{\bibinfo{volume}{72}},
  \bibinfo{pages}{115303}, \doiprefix\url{10.1103/PhysRevB.72.115303}
  (\bibinfo{year}{2005}).

\bibitem{ka:210-niemczyck-natphys-ultrastrong}
\bibinfo{author}{Niemczyk, T.} \emph{et~al.}
\newblock \bibinfo{journal}{\bibinfo{title}{Circuit quantum electrodynamics in
  the ultrastrong-coupling regime}}.
\newblock {\emph{\JournalTitle{Nature Physics}}} \textbf{\bibinfo{volume}{6}},
  \bibinfo{pages}{772–776}, \doiprefix\url{10.1038/nphys1730}
  (\bibinfo{year}{2010}).

\bibitem{ka:210-diazmooji-prl-blochsiegert}
\bibinfo{author}{Forn-D\'{\i}az, P.} \emph{et~al.}
\newblock \bibinfo{journal}{\bibinfo{title}{Observation of the bloch-siegert
  shift in a qubit-oscillator system in the ultrastrong coupling regime}}.
\newblock {\emph{\JournalTitle{Phys. Rev. Lett.}}}
  \textbf{\bibinfo{volume}{105}}, \bibinfo{pages}{237001},
  \doiprefix\url{10.1103/PhysRevLett.105.237001} (\bibinfo{year}{2010}).

\bibitem{ka:217-forndiazlupascu-natphys-usctunableflux}
\bibinfo{author}{Forn-D\'{\i}az, P.} \emph{et~al.}
\newblock \bibinfo{journal}{\bibinfo{title}{Ultrastrong coupling of a single
  artificial atom to an electromagnetic continuum}}.
\newblock {\emph{\JournalTitle{Nat. Phys.}}} \textbf{\bibinfo{volume}{13}},
  \bibinfo{pages}{39} (\bibinfo{year}{2017}).

\bibitem{ka:217-yoshiarasemba-natphys-dscflux}
\bibinfo{author}{Yoshihara, F.} \emph{et~al.}
\newblock \bibinfo{journal}{\bibinfo{title}{Superconducting qubit-oscillator
  circuit beyond the ultrastrong-coupling regime}}.
\newblock {\emph{\JournalTitle{Nat. Phys.}}} \textbf{\bibinfo{volume}{13}},
  \bibinfo{pages}{44} (\bibinfo{year}{2017}).

\bibitem{ka:217-bosmansteele-multimodeusc}
\bibinfo{author}{Bosman, S.} \emph{et~al.}
\newblock \bibinfo{journal}{\bibinfo{title}{Multi-mode ultra-strong coupling in
  circuit quantum electrodynamics}}.
\newblock {\emph{\JournalTitle{npj Quant. Inf.}}} \textbf{\bibinfo{volume}{3}},
  \bibinfo{pages}{46}, \doiprefix\url{10.1038/s41534-017-0046-y}
  (\bibinfo{year}{2017}).
\newblock \bibinfo{note}{ArXiv:1704.06208}.

\bibitem{ka:209-anapparabeltram-prb-ustr}
\bibinfo{author}{Anappara, A.~A.} \emph{et~al.}
\newblock \bibinfo{journal}{\bibinfo{title}{Signatures of the ultrastrong
  light-matter coupling regime}}.
\newblock {\emph{\JournalTitle{Phys. Rev. B}}} \textbf{\bibinfo{volume}{79}},
  \bibinfo{pages}{201303}, \doiprefix\url{10.1103/PhysRevB.79.201303}
  (\bibinfo{year}{2009}).

\bibitem{ka:209-guentertredicucci-nature-switcusc}
\bibinfo{author}{G\"unter, G.} \emph{et~al.}
\newblock \bibinfo{journal}{\bibinfo{title}{Sub-cycle switch-on of ultrastrong
  light–matter interaction}}.
\newblock {\emph{\JournalTitle{Nature}}} \textbf{\bibinfo{volume}{458}},
  \bibinfo{pages}{178--181}, \doiprefix\url{10.1038/nature07838}
  (\bibinfo{year}{2009}).

\bibitem{ka:212-scalari-science-USCTHz}
\bibinfo{author}{Scalari, G.} \emph{et~al.}
\newblock \bibinfo{journal}{\bibinfo{title}{Ultrastrong coupling of the
  cyclotron transition of a 2d electron gas to a thz metamaterial}}.
\newblock {\emph{\JournalTitle{Science}}} \textbf{\bibinfo{volume}{335}},
  \bibinfo{pages}{1323--1326} (\bibinfo{year}{2012}).

\bibitem{ka:217-bayerlange-nanolett-dscsemicond}
\bibinfo{author}{Bayer, A.} \emph{et~al.}
\newblock \bibinfo{journal}{\bibinfo{title}{Terahertz light–matter
  interaction beyond unity coupling strengt}}.
\newblock {\emph{\JournalTitle{Nano Lett.}}} \textbf{\bibinfo{volume}{17}},
  \bibinfo{pages}{6340–6344}, \doiprefix\url{10.1021/acs.nanolett.7b03103}
  (\bibinfo{year}{2017}).

\bibitem{ka:213-stassisavasta-prl-USCSEP}
\bibinfo{author}{Stassi, R.}, \bibinfo{author}{Ridolfo, A.},
  \bibinfo{author}{Di~Stefano, O.}, \bibinfo{author}{Hartmann, M.~J.} \&
  \bibinfo{author}{Savasta, S.}
\newblock \bibinfo{journal}{\bibinfo{title}{Spontaneous conversion from virtual
  to real photons in the ultrastrong-coupling regime}}.
\newblock {\emph{\JournalTitle{Phys. Rev. Lett.}}}
  \textbf{\bibinfo{volume}{110}}, \bibinfo{pages}{243601},
  \doiprefix\url{10.1103/PhysRevLett.110.243601} (\bibinfo{year}{2013}).

\bibitem{ka:217-distefanosavastanori-njp-stimemission}
\bibinfo{author}{Di~Stefano, O.} \emph{et~al.}
\newblock \bibinfo{journal}{\bibinfo{title}{Feynman-diagrams approach to the
  quantum rabi model for ultrastrong cavity qed: stimulated emission and
  reabsorption of virtual particles dressing a physical excitation}}.
\newblock {\emph{\JournalTitle{New J. Phys.}}} \textbf{\bibinfo{volume}{19}},
  \bibinfo{pages}{053010} (\bibinfo{year}{2017}).

\bibitem{ka:214-huanglaw-pra-uscraman}
\bibinfo{author}{Huang, J.-F.} \& \bibinfo{author}{Law, C.~K.}
\newblock \bibinfo{journal}{\bibinfo{title}{Photon emission via vacuum-dressed
  intermediate states under ultrastrong coupling}}.
\newblock {\emph{\JournalTitle{Phys. Rev. A}}} \textbf{\bibinfo{volume}{89}},
  \bibinfo{pages}{033827}, \doiprefix\url{10.1103/PhysRevA.89.033827}
  (\bibinfo{year}{2014}).

\bibitem{ka:212-ridolfo-prl-photonblock}
\bibinfo{author}{Ridolfo, A.}, \bibinfo{author}{Leib, M.},
  \bibinfo{author}{Savasta, S.} \& \bibinfo{author}{Hartmann, M.}
\newblock \bibinfo{journal}{\bibinfo{title}{Photon blockade in the ultrastrong
  coupling regime}}.
\newblock {\emph{\JournalTitle{Phys. Rev. Lett.}}}
  \textbf{\bibinfo{volume}{109}}, \bibinfo{pages}{193602}
  (\bibinfo{year}{2012}).

\bibitem{ka:207-deliberato-prl-uscdynamics}
\bibinfo{author}{De~Liberato, S.}, \bibinfo{author}{Ciuti, C.} \&
  \bibinfo{author}{Carusotto, I.}
\newblock \bibinfo{journal}{\bibinfo{title}{Quantum vacuum radiation spectra
  from a semiconductor microcavity with a time-modulated vacuum rabi
  frequency}}.
\newblock {\emph{\JournalTitle{Phys. Rev. Lett.}}}
  \textbf{\bibinfo{volume}{98}}, \bibinfo{pages}{103602},
  \doiprefix\url{10.1103/PhysRevLett.98.103602} (\bibinfo{year}{2007}).

\bibitem{kr:198-bergmann-rmp-stirap}
\bibinfo{author}{Bergmann, K.}, \bibinfo{author}{Theuer, H.} \&
  \bibinfo{author}{Shore, B.}
\newblock \bibinfo{journal}{\bibinfo{title}{Coherent population transfer among
  quantum states of atoms and molecules}}.
\newblock {\emph{\JournalTitle{Rev. Mod. Phys.}}}
  \textbf{\bibinfo{volume}{70}}, \bibinfo{pages}{1003--1025},
  \doiprefix\url{10.1103/RevModPhys.70.1003} (\bibinfo{year}{1998}).

\bibitem{kr:201-vitanov-advatmolopt}
\bibinfo{author}{Vitanov, N.}, \bibinfo{author}{Fleischhauer, M.},
  \bibinfo{author}{Shore, B.} \& \bibinfo{author}{Bergmann, K.}
\newblock \bibinfo{journal}{\bibinfo{title}{Coherent manipulation of atoms and
  molecules by sequential laser pulses}}.
\newblock {\emph{\JournalTitle{Adv. in At. Mol. and Opt. Phys.}}}
  \textbf{\bibinfo{volume}{46}}, \bibinfo{pages}{55--190}, \doiprefix\url{DOI:
  10.1016/S1049-250X(01)80063-X} (\bibinfo{year}{2001}).

\bibitem{kr:217-vitanovbergmann-rmp}
\bibinfo{author}{Vitanov, N.~V.}, \bibinfo{author}{Rangelov, A.~A.},
  \bibinfo{author}{Shore, B.~W.} \& \bibinfo{author}{Bergmann, K.}
\newblock \bibinfo{journal}{\bibinfo{title}{Stimulated raman adiabatic passage
  in physics, chemistry, and beyond}}.
\newblock {\emph{\JournalTitle{Rev. Mod. Phys.}}}
  \textbf{\bibinfo{volume}{89}}, \bibinfo{pages}{015006},
  \doiprefix\url{10.1103/RevModPhys.89.015006} (\bibinfo{year}{2017}).

\bibitem{ka:210-ashhabnori-pra-uscdyn}
\bibinfo{author}{Ashhab, S.} \& \bibinfo{author}{Nori, F.}
\newblock \bibinfo{journal}{\bibinfo{title}{Qubit-oscillator systems in the
  ultrastrong-coupling regime and their potential for preparing nonclassical
  states}}.
\newblock {\emph{\JournalTitle{Phys. Rev. A}}} \textbf{\bibinfo{volume}{81}},
  \bibinfo{pages}{042311}, \doiprefix\url{10.1103/PhysRevA.81.042311}
  (\bibinfo{year}{2010}).

\bibitem{ka:212-romerosolano-prl-ultrafastgates}
\bibinfo{author}{Romero, G.}, \bibinfo{author}{Ballester, D.},
  \bibinfo{author}{Wang, Y.~M.}, \bibinfo{author}{Scarani, V.} \&
  \bibinfo{author}{Solano, E.}
\newblock \bibinfo{journal}{\bibinfo{title}{Ultrafast quantum gates in circuit
  qed}}.
\newblock {\emph{\JournalTitle{Phys. Rev. Lett.}}}
  \textbf{\bibinfo{volume}{108}}, \bibinfo{pages}{120501},
  \doiprefix\url{10.1103/PhysRevLett.108.120501} (\bibinfo{year}{2012}).

\bibitem{ka:215-felicettisolano-scirept-stateengineernig-USC}
\bibinfo{author}{S.~Felicetti, S.}, \bibinfo{author}{Douce, T.},
  \bibinfo{author}{Romero, G.}, \bibinfo{author}{Milman, P.} \&
  \bibinfo{author}{Solano, E.}
\newblock \bibinfo{journal}{\bibinfo{title}{Parity-dependent state engineering
  and tomography in the ultrastrong coupling regime}}.
\newblock {\emph{\JournalTitle{Sci. Rept.}}} \textbf{\bibinfo{volume}{5}},
  \bibinfo{pages}{11818} (\bibinfo{year}{2015}).

\bibitem{ka:215-kyawsolano-prb-QECandUSC}
\bibinfo{author}{Kyaw, T.~H.}, \bibinfo{author}{Herrera-Mart\'{\i}, D.~A.},
  \bibinfo{author}{Solano, E.}, \bibinfo{author}{Romero, G.} \&
  \bibinfo{author}{Kwek, L.-C.}
\newblock \bibinfo{journal}{\bibinfo{title}{Creation of quantum error
  correcting codes in the ultrastrong coupling regime}}.
\newblock {\emph{\JournalTitle{Phys. Rev. B}}} \textbf{\bibinfo{volume}{91}},
  \bibinfo{pages}{064503}, \doiprefix\url{10.1103/PhysRevB.91.064503}
  (\bibinfo{year}{2015}).

\bibitem{ka:217-andersenblais-njp-transmonUSC}
\bibinfo{author}{Andersen, C.} \& \bibinfo{author}{Blais, A.}
\newblock \bibinfo{journal}{\bibinfo{title}{Ultrastrong coupling dynamics with
  a transmon qubit}}.
\newblock {\emph{\JournalTitle{New J. Phys.}}} \textbf{\bibinfo{volume}{19}},
  \bibinfo{pages}{023022} (\bibinfo{year}{2017}).

\bibitem{ka:213-ashab-pra-superradiance}
\bibinfo{journal}{\bibinfo{author}{Ashhab, S.}}
\newblock {\emph{\JournalTitle{Phys. Rev. A}}} \textbf{\bibinfo{volume}{87}},
  \bibinfo{pages}{013826} (\bibinfo{year}{2013}).

\bibitem{ka:216-jaakorabl-pra-beyonddicke}
\bibinfo{author}{Jaako, T.}, \bibinfo{author}{Xiang, Z.-L.},
  \bibinfo{author}{Garcia-Ripoll, J.~J.} \& \bibinfo{author}{Rabl, P.}
\newblock \bibinfo{journal}{\bibinfo{title}{Ultrastrong-coupling phenomena
  beyond the dicke model}}.
\newblock {\emph{\JournalTitle{Phys. Rev. A}}} \textbf{\bibinfo{volume}{94}},
  \bibinfo{pages}{033850}, \doiprefix\url{10.1103/PhysRevA.94.033850}
  (\bibinfo{year}{2016}).

\bibitem{km:supplemental}
\bibinfo{title}{See the section Methods for further details on Rabi model, STIRAP and control of dynamical Stark  shift}).

\bibitem{ka:206-siebrafalci-optcomm-stirap}
\bibinfo{author}{Siewert, J.}, \bibinfo{author}{Brandes, T.} \&
  \bibinfo{author}{Falci, G.}
\newblock \bibinfo{journal}{\bibinfo{title}{Adiabatic passage with
  superconducting nanocircuits}}.
\newblock {\emph{\JournalTitle{Opt Commun}}} \textbf{\bibinfo{volume}{264}},
  \bibinfo{pages}{435 -- 440},
  \doiprefix\url{http://dx.doi.org/10.1016/j.optcom.2005.12.083}
  (\bibinfo{year}{2006}).
\newblock \bibinfo{note}{Quantum Control of Light and Matter. In honor of the
  70th birthday of Bruce Shore}.

\bibitem{ka:208-weinori-prl-stirapqcomp}
\bibinfo{author}{Wei, L.~F.}, \bibinfo{author}{Johansson, J.~R.},
  \bibinfo{author}{Cen, L.~X.}, \bibinfo{author}{Ashhab, S.} \&
  \bibinfo{author}{Nori, F.}
\newblock \bibinfo{journal}{\bibinfo{title}{Controllable coherent population
  transfers in superconducting qubits for quantum computing}}.
\newblock {\emph{\JournalTitle{Phys. Rev. Lett.}}}
  \textbf{\bibinfo{volume}{100}}, \bibinfo{pages}{113601},
  \doiprefix\url{10.1103/PhysRevLett.100.113601} (\bibinfo{year}{2008}).

\bibitem{ka:209-siebrafalci-prb}
\bibinfo{author}{Siewert, J.}, \bibinfo{author}{Brandes, T.} \&
  \bibinfo{author}{Falci, G.}
\newblock \bibinfo{journal}{\bibinfo{title}{Advanced control with a cooper-pair
  box: Stimulated raman adiabatic passage and fock-state generation in a
  nanomechanical resonator}}.
\newblock {\emph{\JournalTitle{Phys. Rev. B}}} \textbf{\bibinfo{volume}{79}},
  \bibinfo{pages}{024504}, \doiprefix\url{10.1103/PhysRevB.79.024504}
  (\bibinfo{year}{2009}).

\bibitem{ka:216-kumarparaoanu-natcomm-stirap}
\bibinfo{author}{Kumar, K.}, \bibinfo{author}{Veps\"al\"ainen, A.},
  \bibinfo{author}{Danilin, S.} \& \bibinfo{author}{Paraoanu, G.}
\newblock \bibinfo{journal}{\bibinfo{title}{Stimulated raman adiabatic passage
  in a three-level superconducting circuit}}.
\newblock {\emph{\JournalTitle{Nat. Comm.}}} \textbf{\bibinfo{volume}{7}},
  \bibinfo{pages}{10628}, \doiprefix\url{10.1038/ncomms10628}
  (\bibinfo{year}{2016}).

\bibitem{ka:216-xuhanzhao-natcomm-ladderstirap}
\bibinfo{author}{Xu, H.} \emph{et~al.}
\newblock \bibinfo{journal}{\bibinfo{title}{Coherent population transfer
  between uncoupled or weakly coupled states in ladder-type superconducting
  qutrits}}.
\newblock {\emph{\JournalTitle{Nature Communications}}}
  \textbf{\bibinfo{volume}{7}}, \bibinfo{pages}{11018},
  \doiprefix\url{doi:10.1038/ncomms11018} (\bibinfo{year}{2016}).

\bibitem{ka:216-vepsalainen-photonics-squtrit}
\bibinfo{author}{Veps\"al\"ainen, A.}, \bibinfo{author}{Danilin, S.},
  \bibinfo{author}{Paladino, E.}, \bibinfo{author}{Falci, G.} \&
  \bibinfo{author}{Paraoanu, G.~S.}
\newblock \bibinfo{journal}{\bibinfo{title}{Quantum control in qutrit systems
  using hybrid rabi-stirap pulses}}.
\newblock {\emph{\JournalTitle{Photonics}}} \textbf{\bibinfo{volume}{3}},
  \bibinfo{pages}{62}, \doiprefix\url{10.3390/photonics3040062}
  (\bibinfo{year}{2016}).

\bibitem{ka:217-falci-fortphys-fqmt}
\bibinfo{author}{Falci, G.} \emph{et~al.}
\newblock \bibinfo{journal}{\bibinfo{title}{Advances in quantum control of
  three-level superconducting circuit architectures}}.
\newblock {\emph{\JournalTitle{Fort. Phys.}}} \textbf{\bibinfo{volume}{65}},
  \bibinfo{pages}{1600077}, \doiprefix\url{10.1002/prop.201600077}
  (\bibinfo{year}{2017}).

\bibitem{ka:216-distefano-pra-twoplusone}
\bibinfo{author}{Di~Stefano, P.~G.}, \bibinfo{author}{Paladino, E.},
  \bibinfo{author}{Pope, T.~J.} \& \bibinfo{author}{Falci, G.}
\newblock \bibinfo{journal}{\bibinfo{title}{Coherent manipulation of
  noise-protected superconducting artificial atoms in the lambda scheme}}.
\newblock {\emph{\JournalTitle{Phys. Rev. A}}} \textbf{\bibinfo{volume}{93}},
  \bibinfo{pages}{051801}, \doiprefix\url{10.1103/PhysRevA.93.051801}
  (\bibinfo{year}{2016}).

\bibitem{ka:212-rigettisteffen-prb-trasmonshapphire}
\bibinfo{author}{Rigetti, C.} \emph{et~al.}
\newblock \bibinfo{journal}{\bibinfo{title}{Superconducting qubit in a
  waveguide cavity with a coherence time approaching 0.1 ms}}.
\newblock {\emph{\JournalTitle{Phys. Rev. B}}} \textbf{\bibinfo{volume}{86}},
  \bibinfo{pages}{100506}, \doiprefix\url{10.1103/PhysRevB.86.100506}
  (\bibinfo{year}{2012}).

\bibitem{kr:214-paladino-rmp}
\bibinfo{author}{Paladino, E.}, \bibinfo{author}{Galperin, Y.},
  \bibinfo{author}{Falci, G.} \& \bibinfo{author}{Altshuler, B.}
\newblock \bibinfo{journal}{\bibinfo{title}{1/f noise: implications for
  solid-state quantum information}}.
\newblock {\emph{\JournalTitle{Rev. Mod. Phys.}}}
  \textbf{\bibinfo{volume}{86}}, \bibinfo{pages}{361--418},
  \doiprefix\url{10.1103/RevModPhys.86.361} (\bibinfo{year}{2014}).

\bibitem{ka:203-paladino-advssp-decoherence}
\bibinfo{author}{Paladino, E.}, \bibinfo{author}{Faoro, L.} \&
  \bibinfo{author}{Falci, G.}
\newblock \bibinfo{title}{Decoherence due to discrete noise in josephson
  qubits}.
\newblock In \bibinfo{editor}{{Kramer, B}} (ed.) \emph{\bibinfo{booktitle}{Adv.
  in Sol. State Phys.}}, vol.~\bibinfo{volume}{{\bf 43}} of
  \emph{\bibinfo{series}{{ADVANCES IN SOLID STATE PHYSICS}}},
  \bibinfo{pages}{747--762} (\bibinfo{organization}{{Deutsch Phys Gesell,
  Arbeitskreis Festkorperphys}}, \bibinfo{year}{2003}).
\newblock \bibinfo{note}{Spring Meeting of the Arbeitskreis-Festkorperphysik of
  the Deutsche-Physikalische-Gesellschaft, DRESDEN, GERMANY, MAR 24-28, 2003}.

\bibitem{ka:215-peterergustaffson-prl-transmexclevel}
\bibinfo{author}{Peterer, M.~J.} \emph{et~al.}
\newblock \bibinfo{journal}{\bibinfo{title}{Coherence and decay of higher
  energy levels of a superconducting transmon qubit}}.
\newblock {\emph{\JournalTitle{Phys. Rev. Lett.}}}
  \textbf{\bibinfo{volume}{114}}, \bibinfo{pages}{010501},
  \doiprefix\url{10.1103/PhysRevLett.114.010501} (\bibinfo{year}{2015}).

\bibitem{ka:213-falci-prb-stirapcpb}
\bibinfo{author}{Falci, G.} \emph{et~al.}
\newblock \bibinfo{journal}{\bibinfo{title}{Design of a lambda system for
  population transfer in superconducting nanocircuits}}.
\newblock {\emph{\JournalTitle{Phys. Rev. B}}} \textbf{\bibinfo{volume}{87}},
  \bibinfo{pages}{214515--1,214515--13},
  \doiprefix\url{10.1103/PhysRevB.87.214515} (\bibinfo{year}{2013}).

\bibitem{ka:218-ridolfofalci-epj-usc}
\bibinfo{journal}{\bibinfo{author}{Ridolfo, A.}, \bibinfo{author}{Falci, G.},
  \bibinfo{author}{Pellegrino, F.} \& \bibinfo{author}{Paladino, E.}}
\newblock {\emph{\JournalTitle{arXiv 1805.07079}}}).

\bibitem{ka:215-distefano-prb-cstirap}
\bibinfo{author}{Di~Stefano, P.~G.}, \bibinfo{author}{Paladino, E.},
  \bibinfo{author}{D'Arrigo, A.} \& \bibinfo{author}{Falci, G.}
\newblock \bibinfo{journal}{\bibinfo{title}{Population transfer in a lambda
  system induced by detunings}}.
\newblock {\emph{\JournalTitle{Phys. Rev. B}}} \textbf{\bibinfo{volume}{91}},
  \bibinfo{pages}{224506}, \doiprefix\url{10.1103/PhysRevB.91.224506}
  (\bibinfo{year}{2015}).

\bibitem{kr:210-singercalarco-rmp-trapionsnumerics}
\bibinfo{author}{Singer, K.} \emph{et~al.}
\newblock \bibinfo{journal}{\bibinfo{title}{Colloquium: Trapped ions as quantum
  bits: Essential numerical tools}}.
\newblock {\emph{\JournalTitle{Rev. Mod. Phys.}}}
  \textbf{\bibinfo{volume}{82}}, \bibinfo{pages}{2609--2632},
  \doiprefix\url{10.1103/RevModPhys.82.2609} (\bibinfo{year}{2010}).

\end{thebibliography}

\section*{Discussion}
In conclusion, we propose the dynamical detection of the USC regime of light-matter interaction 
using an ancillary atomic level as a probe. The opening of a USC-specific channel for population transfer is witnessed by detecting two-photons in the harmonic mode. 
This process, coherently amplified by STIRAP, marks the symmetry broken by the USC 
counterrotating $g$ term in $H_R$ Eq.(\ref{eq:rabiH2}), which determines the violation of the conservation of $N$ 
Relying on coherent dynamics, the experiment we propose would be a benchmark for adiabatic quantum control of circuit QED architectures~\cite{ka:217-falci-fortphys-fqmt,ka:215-distefano-prb-cstirap} in the USC regime. STIRAP is known to be superior to other protocols~\cite{kr:198-bergmann-rmp-stirap,kr:201-vitanov-advatmolopt,kr:217-vitanovbergmann-rmp} in the $\Lambda$ scheme, as SEP~\cite{ka:213-stassisavasta-prl-USCSEP} or Raman oscillations~\cite{ka:214-huanglaw-pra-uscraman}. What makes it unique is the possibility to operate in $V$-configuration, which is resilient to the presence of stray AA-mode couplings, inevitable in three-level USC solid state architectures.
Flux qubits, offering the largest $g/\omega_c > 1$ fabricated so far, also meet all the quantum hardware requirements for the experiment.

\section*{Methods}
\subsection{The Jaynes-Cummings and the Rabi models}
\label{suppl:rabispec}
The JC Hamiltonian describes a two-level atom coupled to an electromagnetic mode in the RWA
\begin{equation}
\label{eq:JCmodel}
H_{JC} = \varepsilon \,\ketbra{e}{e} + 
\omega_c \,a^\dagger a + g \,\big[a \,\ketbra{e}{g}+ a^\dagger \,\ketbra{g}{e} \big]  
\end{equation}
where with no loss of generality we assume $g>0$. The ground state is factorized
$\ket{\phi_0}= \ket{0 g}$ with ${\cal E}_0 = 0$ whereas the rest of the spectrum is arranged in doublets,
$\ket{\phi_{N \sigma}}$, with fixed number of excitations $N \leftrightarrow a^\dagger a + \ketbra{e}{e}$ and labeled by the extra quantum number 
$\sigma=\pm$. At resonance, $\varepsilon = \omega_c$, eigenstates/eigenvalues of $H_{JC}$ in the atom-mode product basis are given by
$$
\ket{\phi_{N \mp}} = \frac{\ket{N-1, e} \mp \ket{N, g}}{\sqrt{2}} \quad; \quad {\cal E}_{N\mp}= N \omega_c \mp \sqrt{N} \, g
$$
Eigenstates of $H_R$, Eq.~(1) in the main text, do not have a well defined $N$ only 
the parity of $N$ being conserved, $\sum_n (-1)^n \ketbra{n}{n} \,[\ketbra{g}{g}-\ketbra{e}{e}]$.
Many of the amplitudes in the decomposition in product states (reported in the text)
vanish. For instance for the dressed ground state 
$\ket{0g} \to \ket{\Phi_0}$ all 
$c_{0\,n}:= \braket{n\,g}{\Phi_0}$ 
($d_{0\,n}:= \braket{n\,e}{\Phi_0}$) with odd (even) 
number of photons $n$ vanish
$$
\ket{\Phi_0} = \sum_{m=0}^\infty \big[c_{0\,2m} \ket{2m \,g} +  d_{0\,2m+1} \ket{2m+1 \,e} \big]
$$
For not too large $g$ eigenstates  $\ket{\Phi_j}$ of the $H_R$, Eq.~(1) in the main text, can be enumerated with the same quantum numbers of the JC limit, $j \equiv (N, \mp)$. 
For them conservation of the parity  of $N$ implies the following structure 
$$
\ket{\Phi_{N \mp}} = \sum_{m> -N/2}  \big[c_{N\mp,\,N+2m} \ket{N+2m, \,g} 
+  d_{N\mp,\,N+2m-1} \ket{N+2m-1, \,e}\big]
$$
Some of the essential features of STIRAP in the USC regime emerge already treating the counterrotating term 
of $H_R$, Eq.~(1) in the main text, in perturbation theory. To this end it is convenient to generalize slightly 
$H_R$, Eq.~(1) in the main text, 
by allowing for a different coupling constant $g_c \neq g$ for the counterrotating term.
The leading corrections of interest to the JC ground state $\ket{0g}$ are 
\begin{equation}
\label{eq:rabi-tls-perturbative}
\begin{aligned}
c_{00} &= \braket{0\,g}{\Phi_0} = 
\frac{4 \omega_c^2 - 2 g^2}{\sqrt{(4 \omega_c^2 - 2 g^2)^2 + g_c^2 (4 \omega_c^2 + 2 g^2)}}
\\
c_{02} &=\braket{2\,g}{\Phi_0} =  g_c \frac{ \sqrt{2} g}{4 \omega_c^2 - 2 g^2}
\end{aligned} 
\end{equation}
Notice that these expressions are perturbative in $g_c$ but nonperturbative in $g$, the JC model being recovered for $g_c \to 0$. 
The nonzero overlap with the $N>0$ states marks the emergence of USC.  
At leading order $c_{02} \propto g_c g$ therefore in the physical case $g_c=g$ the amplitude 
$c_{02} \propto g^2$ (see Fig.~1(b) in the main text). The fact that STIRAP depends on $\Omega_s \propto |c_{02}|$ 
while SEP depends on $|c_{02}|^2 \propto g^4$, which is much smaller, is one of the assets of coherent amplification.

The first JC doublet $\ket{\phi_{1 \mp}} \to \ket{\Phi_{1 \mp}}$ enters $V$-STIRAP. 
In this case $\braket{0\,e}{\Phi_{1 \mp}} \approx 1/\sqrt{2} \approx \mp \braket{1\,g}{\Phi_{1 \mp}}$ 
the relevant corrections in leading order being  
$$
d_{1\mp,2} = \braket{2\,e}{\Phi_{1 \mp}} = - g_c\,
\frac{2 \omega_c \pm g}{(2 \omega_c \pm g)^2 - 3 g^2}
$$
The nonzero overlap of $\ket{\Phi_{1 \mp}}$ with the $N > 1$ states marks the emergence of USC. It is worth stressing that the efficiency of V-STIRAP depends on $\Omega_s \propto |c_{1 \pm, 2}| \propto g_c$, i.e. 
for relatively small $g$ is much more efficient than $\Lambda-$STIRAP. 

\begin{figure}[t!]
\centering
\resizebox{0.7\columnwidth}{!}{\includegraphics{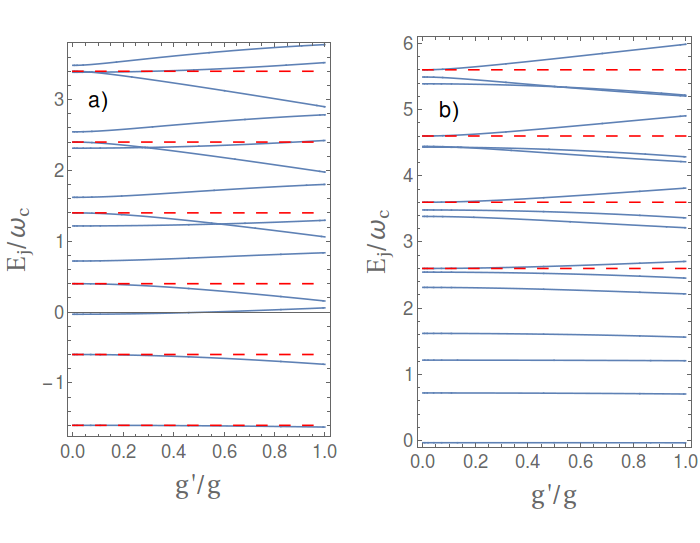}}
\caption{(color online) Spectrum of the three-level system coupled in ladder configuration with an harmonic mode, in the 
USC regime, $g=0.25 \,\omega_c$, with one mode at resonance $\epsilon = \omega_c$, 
as a function of the relative strength of the stray coupling $\eta=g^\prime/g$. 
Red dashed lines are the energies of the uncoupled ancillary states $\ket{n u}$, i.e. when $g^\prime=0$. 
(a) Spectrum for the level configuration used for the $\Lambda$-scheme 
Eqs.~(\ref{eq:H-threelevel-lambda-JC},\ref{eq:H-threelevel-lambda-count}), with $\epsilon_p=1.6 \,\omega_c$.
(b) Spectrum for the level configuration used for the $V$-scheme 
Eqs.~(\ref{eq:H-threelevel-vee-JC},\ref{eq:H-threelevel-vee-count}), with $\epsilon_p=1.6 \,\omega_c$.
\label{fig:3LSspectrum}}
\end{figure}

\subsection{Lambda STIRAP in the USC regime}
In the simplest instance $\Lambda$-STIRAP is obtained by adding an uncoupled atomic level $\ket{u}$ to $H_R$. 
In the main text we introduced the resulting Hamiltonian 
\begin{equation}\label{eq:rabiH3}
\begin{aligned}
H_0 = -\varepsilon^\prime \,\ketbra{u}{u} + H_R + \omega_c \,a^\dagger a \otimes \ketbra{u}{u}.
\end{aligned}\end{equation}
where the last term completes the identity operator for the AA in 
passing from the two-level system in $H_R$ to the three-level system in $H_0$.  
We take $\varepsilon^\prime \gg \varepsilon = \omega_c \gtrsim g,g_c$.  
The control field is coupled to the AA, and we consider a  
two tone $W(t)$ with frequencies 
$\omega_p = E_0 + \varepsilon^\prime + \delta_p$ and $\omega_s = E_0 + \varepsilon^\prime - 2 \omega_c+ \delta_s$, 
and for simplicity $\delta_p = \delta_s =0$ in most simulations. 
For large $\varepsilon^\prime$ the drive couples only with the 
$u\!-\!g$ transition, yielding Eq.~(3) in the main text. If $g$ is not too large we can neglect terms with $n>2$, which are detuned and have smaller amplitude $c_{0n}(g)$.  
The simple $\Lambda$ configuration~\cite{kr:201-vitanov-advatmolopt} $H_C^{\Lambda}(t)$ 
is obtained by retaining the resonant and corotating parts of $W(t)$.
In order to operate STIRAP we take slowly varying envelopes with Gaussian shape 
$$
\mathscr{W}_s(t) = \bar{\mathscr{W}}_s \,\mathrm{e}^{- [(t+\tau)/T]^2}
\quad ; \quad \mathscr{W}_p(t) = \bar{\mathscr{W}}_p \,\mathrm{e}^{- [(t-\tau)/T]^2}
$$   
where the delay $\tau>0$ implements the ``counterintuitive'' sequence. 

\subsubsection{General properties and optimization}
STIRAP relies on resonant external fields inducing destructive interference. Faithful and selective
coherent population transfer is achieved by adiabatic dynamics~\cite{kr:198-bergmann-rmp-stirap,kr:217-vitanovbergmann-rmp}. Adiabaticity requires sufficiently large pulse amplitudes, satisfying the so called ''global condition``~\cite{kr:201-vitanov-advatmolopt} 
$$
c_{00} \bar{\mathscr{W}}_p , c_{02} \bar{\mathscr{W}}_s > 10 \, T
$$
Besides the efficiency the virtue of STIRAP is the remarkable robustness against variation of the parameters. Indeed STIRAP is not very sensitive to slight deviations from optimal values of the parameters (pulse shapes and amplitudes, detunings and delay), thus efficient amplification of the 
$\ket{0u} \to \ket{2u}$ channel does not require fine tuning of too many parameters. In the simulations we used the standard figures $\tau=0.75\,T$ and $\delta_s = \delta_p =0$ for conventional STIRAP and since it is known that the best robustness is obtained for equal peak Rabi frequencies~\cite{kr:201-vitanov-advatmolopt}, 
$\max_t[\Omega_s(t)]=\max_t[\Omega_p(t)]$, 
we considered an attenuated pump field 
$$\bar{\mathscr{W}_p} = \kappa_p ^{\Lambda}\bar{\mathscr{W}_s} \quad, \quad 
\kappa_p^{\Lambda} = c_{02}(g)/c_{00}(g)$$ 
In this situation the only relevant sensitivity to take care of is related to deviations 
from the two-photon resonance condition $\delta=0$, where 
$\delta:=\delta_s - \delta_p = 2 \omega_c - (\omega_p-\omega_s)$ is the two-photon detuning.  
For $\delta \neq 0$ no exact dark state exists and no adiabatic pattern connects the initial and the
target state, but if $|\delta| \lesssim \max_t[\Omega_k(t)]/5$ efficient population transfer still occurs via 
diabatic processes~\cite{kr:201-vitanov-advatmolopt}. These considerations apply to both $\Lambda$ and $V$ STIRAP.

\subsubsection{Dynamical Stark shift compensation}
The simple standard form of the control Hamiltonian in $\Lambda$ configuration, $H_C^\Lambda(t)$, reported in the text 
is obtained assuming that field amplitudes $\mathscr{W}_k$ are small enough to be negligible except for fields quasi-resonant 
and corotating. This may not be the case since for small $g$ enforcing adiabaticity requires a large Stokes field $\mathscr{W}_s$. 
In Fig.~2 in the main text  we used a peak value of $\mathscr{W}_s$ which would yield
$e\!-\!g$ Rabi oscillations with angular frequency $\Omega_0=600\,\mathrm{MHz}$). Large fields produce Stark shifts of the detuned transitions: shift  of level $j$ due to the coupling to level $i$ under the action of the $k$ field is given by
$$
S_{ij}^{(k)}(t)=\left
\lvert \frac{\eta_{ij}{\mathscr W}_{k}(t)}{2} \right\rvert^2 \left( \frac{1}{E_{i}
-E_j -\omega_{k}} +
\frac{1}{
E_{i}
-E_j + \omega_{k}} \right)
$$ 
where $\eta_{ij}$ is the ratio of the \enquote{dipole} matrix element of the selected transition with the 
reference $e-g$ one. The main effect is due to 
the term $W_s(t) [\ketbra{0u}{\Phi_0}+ \ketbra{\Phi_0}{0u}]$ and affects the $\ket{0u}-\ket{\Phi_0}$ transition. In the three-level approximation
the dynamical Stark shift induces a stray detunings
$\delta(t)= - S_{0u, 0}^{(s)}(t)$, comparable to $\Omega_s$, which completely suppresses 
STIRAP~\cite{ka:217-falci-fortphys-fqmt,ka:218-ridolfofalci-epj-usc}. 

Actually the three-level analysis must be generalized to account for the multilevel nature of the system.
The relevant stray detuning is  $\delta(t)= \sum_{j\neq 2u} S_{2u, j}^{(s)}(t) - \sum_{j\neq 0u} S_{0u, j}^{(s)}(t)$, this structure determining self-compensations which fortunately mitigate the detrimental effect 
of dynamical Stark shifts. We studied numerically this problem considering up to 40 levels and a control field with the structure 
$H_C(t)= W(t) [(\ketbra{u}{g} + (1/\eta) \ketbra{g}{e}) + \mbox{h.c.} ]$, which  
describes the experimentally relevant case of a ladder type ``dipole`` coupling to the AA,  
$\eta$ being the ratio between the corresponding matrix elements
(Fig.~2 in the main text).

The full signal can be recovered if appropriately crafted control is used as shown in Fig.~2 in the main text (thin red curve). 
One option is to use a phase modulation of the Stokes pulse, as explained in~\cite{ka:216-distefano-pra-twoplusone}, designed to compensate the effect of the Stokes field coupled to the $u\!-\!g$ transition only.
The fact that this is the relevant source of noise is suggested by success of the strategy even when the 
drive fully couples to the AA Ladder. 
Another option, namely to add a suitably designed off-resonant tone $W(t) \to W(t)+\mathscr{W}_s(t) \,\cos (2 \omega_s t)$, will be discussed elsewhere. It is worth stressing that both advanced control methods are designed on the basis of a qualitative analysis of the system's Hamiltonian, therefore they can be further refined using Optimal Control Theory~\cite{kr:210-singercalarco-rmp-trapionsnumerics}.
    
Of course by increasing the coupling $g$ a large $\mathscr{W}_s$ is not needed any more and no stray detuning is induced. Dynamical Stark shifts can be neglected and naturally disappear for $g \gtrsim 0.4$.

\section*{Acknowledgements}
We acknowledge discussions with F. Deppe, S. Savasta, R. Fazio, M. Paternostro, M. Hartmann, S. Paraoanu, A. Wallraff, J. Koch, S. De Liberato, C. Ciuti, A. Tomadin. 

\section*{Author contributions statement}
G.F. and E.P. conceived the model; A.R. carried out the numerical calculations; G.F. carried out the analytical calculations; G.F.,A.R. and G.P.D. analyzed the data. All authors discussed the physics and contributed to writing 
the manuscript.

\section*{Competing interests}
The authors declare no competing interests.


\end{document}